\documentclass[a4paper,11pt]{article}
\pdfoutput=1 

\usepackage{jinstpub} 
\usepackage[utf8]{inputenc}

\usepackage[T1]{fontenc}
\usepackage{graphicx}
\usepackage{dcolumn}
\usepackage{bm}

\usepackage{color}

\title{Operation and performance of a dual-phase crystalline/vapor xenon time projection chamber}

\author[a]{S. Kravitz,}\emailAdd{swkravitz@lbl.gov}
\author[a]{H. Chen,}
\author[a,b]{R. Gibbons,}
\author[a]{S.J. Haselschwardt,}
\author[a]{Q. Xia,} 
\author[a]{and P. Sorensen}\emailAdd{pfsorensen@lbl.gov}
\affiliation[a]{Lawrence Berkeley National Laboratory, 1 Cyclotron Road, Berkeley, CA 94720, USA}
\affiliation[b]{University of California, Berkeley, Department of Physics, 366 Physics North MC 7300, Berkeley, CA 94720, USA}

\date{\today}

\abstract{We have built and operated a crystalline/vapor xenon TPC, with the goal of improving searches for dark matter. The motivation for this instrument is the fact that beta decays from the radon decay chain to the ground state presently limit the state-of-the-art liquid/vapor xenon experiments. In contrast, a crystalline xenon target has the potential to exclude, or tag and reject radon-chain backgrounds.  As a preamble to demonstrating such capabilities, the present article makes a first demonstration of the operation of a crystalline/vapor xenon TPC with electroluminescence (gas gain) for the electron signal readout. It also shows that the scintillation yield in crystalline xenon appears to be identical to that in liquid xenon, in contrast to previous results.}

\begin{document}
\maketitle
\flushbottom

\section{Introduction}
\label{sec:intro}
A new generation of massive dark matter direct detection experiments will soon begin operation. The state-of-the-art detector technology is a liquid/vapor xenon time projection chamber (TPC), with experiments such as LZ~\cite{LZ:2019sgr} and XENONnT~\cite{XENON:2020kmp} expected to search a significant new region of dark matter parameter space. However, they will be limited by radon backgrounds rather than the irreducible neutrino flux from solar, atmospheric and diffuse cosmic sources~\cite{Billard:2013qya,OHare:2021utq}. Whatever the experimental outcome -- improved exclusion limits, a few tantalizing events, or a dark matter signal detection -- there will be strong motivation to reach the irreducible neutrino detection limit~\cite{Bottaro:2021snn}. Removing radon-related backgrounds would result in an experiment with neutrinos as the dominant background. 

We have built a small crystalline/vapor xenon TPC in order to develop the instrumentation to achieve this goal. Such an instrument is expected to offer two significant new advantages: (1) exclusion of radon which emanates from detector surfaces, and (2) full tagging of radon or radon daughter nuclei which decay in the active xenon target. These advantages are uniquely available in a crystalline state, in which atoms are localized at a fixed point in the crystal lattice. Coupled with the well-known time structure of the radon decay chains, a unique five-dimensional $(x,y,z,E,t)$ signature of radon decays allows this background to be identified and removed. This signature vanishes in the liquid state due to convective fluid flow, which destroys the spatial correlation between radon decays and their daughter isotopes.

In addition to these two advantages, the crystalline state offers two further benefits compared with liquid. First, the crystal possesses an approximately 17\%~\cite{NIST, Eatwell:1961} higher density for increased particle stopping power and target mass. Second, an increase in electron mobility of about $\times1.6$ \cite{Yoo:2014dca}, which can aid in the rejection of pileup and accidental coincidence backgrounds.

All of the key operational properties of the liquid/vapor xenon TPC are expected to be preserved in the crystalline state, though not all of these have been demonstrated. In this article we show that a crystalline xenon dual-phase TPC behaves essentially in the same fashion as the well-known liquid xenon dual-phase TPC. Additionally, we report a new measurement of the scintillation yield of crystalline xenon relative to liquid xenon, which we find to be identical. In contrast, previous work~\cite{Yoo:2014dca} found a 15\% reduction in scintillation yield in the crystalline state. The present results align with our expectation due to the similarity of the band gap: 9.22~eV in liquid compared to 9.27~eV in crystal~\cite{Asaf:1974}. We surmise that the previously measured reduction may be attributed to losses at the photomultiplier window, as a result of a portion of the scintillation from crystalline xenon shifting to a lower wavelength.

\section{Instrument}
\label{sec:instru}

\begin{figure}
\centering
\includegraphics[width=.49\textwidth,angle=0]{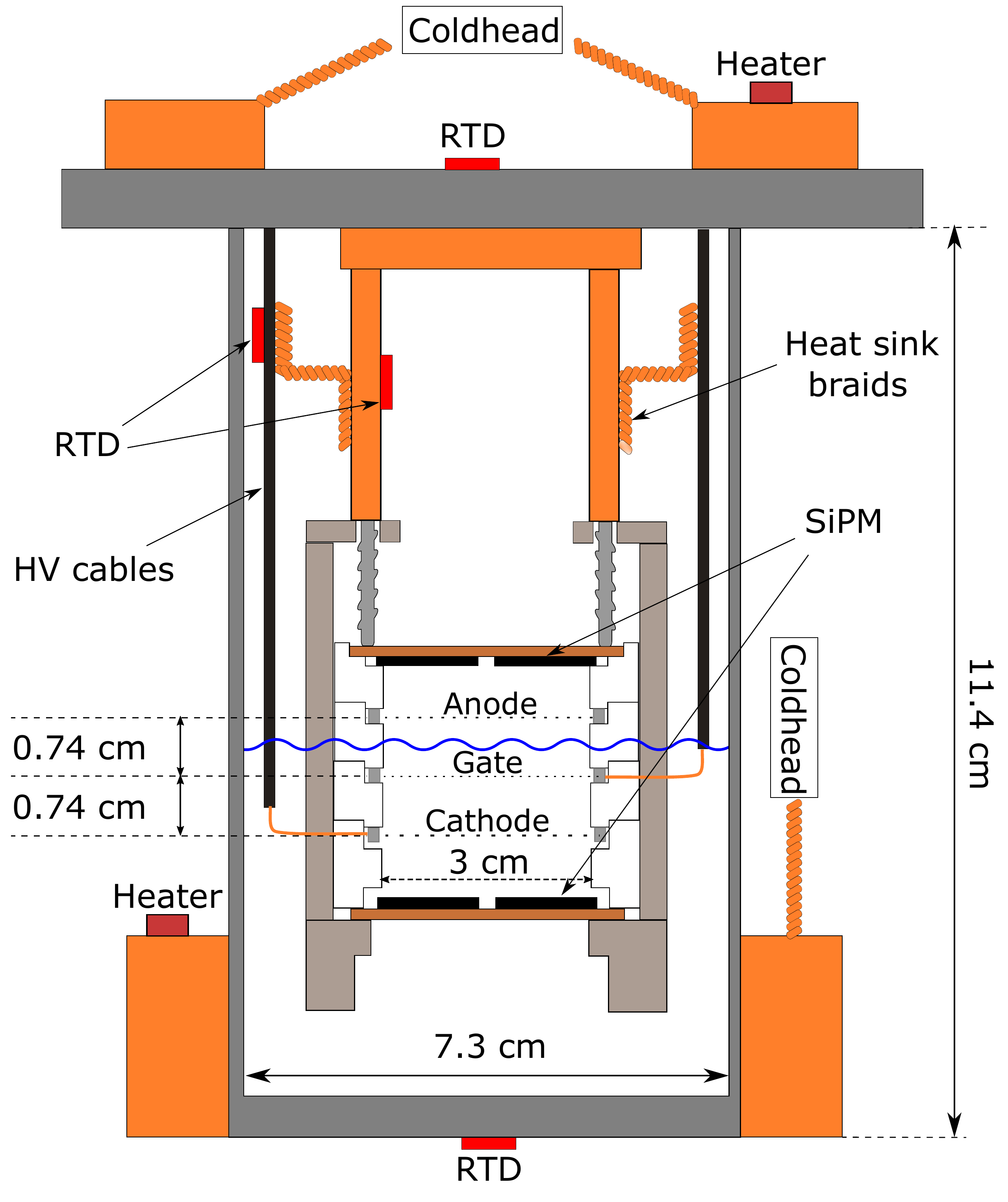}
\caption{\label{fig:schematic} Schematic cross-sectional view of the inner cryostat and TPC. The interface between vapor and liquid or crystal xenon is indicated by the wavy line between gate and anode. Various copper thermal connections are indicated in orange.}
\end{figure}

A schematic of our apparatus is shown in Fig.~\ref{fig:schematic}. The TPC at its core consists of gate, cathode, and anode electrodes housed in reflective polytetrafluoroethylene (PTFE) holders, with arrays of silicon photomultipliers (SiPMs) at the top and bottom to detect light from interactions in the TPC. During operation, the detector is filled with either liquid or crystal xenon to a level between the gate and anode grid, with the region above staying fully in the vapor xenon phase (this requires cooling below the triple point for crystal/vapor phase operation). 

\subsection{Particle detection with a dual phase xenon TPC}
In a traditional liquid-xenon TPC, interactions in the condensed xenon phase, such as gamma-ray Compton scattering or photoabsorption on atomic electrons (electron recoils or ERs) or neutron scattering off xenon nuclei (nuclear recoils or NRs), excite and ionize the xenon producing scintillation light and free electrons. The scintillation light is promptly detected by photodetectors such as SiPMs (primarily in the bottom array due to reflection at the boundary with the vapor phase) - this signal is referred to as the S1 - while the electrons drift toward the anode in the presence of an electric field set by the voltage difference between the cathode and gate grids. 

Drifting electrons that are extracted to the vapor phase then produce secondary electroluminescence light - the S2 signal - due to energetic collisions with the xenon vapor in the much higher field between the gate and anode. The time delay between the S1 and S2 can be used to infer the position of the interaction along the drift direction ($z$); the distribution of S2 light in the top photodetector array can in principle be used to determine the transverse ($x$ and $y$) position of the interaction. The size of the S1 and S2 signals provides a measure of the deposited energy of the interaction.

\subsection{TPC design}
The TPC in our apparatus is shown in Fig.~\ref{fig:schematic} and has an active region diameter of 3.0~cm, set by the inner diameter of the PTFE. The drift region between the gate and cathode is 0.74~cm tall, as is the extraction region between the gate and anode. The gate, anode, and cathode electrodes are each formed by a set of parallel wire grids. The stainless steel grid wires\footnote{\href{https://calfinewire.com/products/wire-drawing/}{California Fine Wire, “ultra finish”}} have 100~$\mu$m diameter, and are strung on stainless steel frames with a 2~mm pitch. The anode potential is always kept at ground, while the gate is typically kept at -2.4~kV and the cathode varies in the $[-2.6, -3.2]$~kV range for field-on measurements (grounded for field-off measurements). 

\subsection{Electronics and data acquisition}
Two arrays (one above the anode, one below the cathode) of four SiPMs\footnote{\href{https://www.hamamatsu.com/us/en/product/optical-sensors/mppc/mppc_mppc-array/index.html}{Hamamatsu VUV-4 S13370}} form the photon detection planes. Each array is mounted on copper-plated G10 circuit boards, with bias and signal readout connections via coaxial cables\footnote{\href{https://www.pasternack.com/flexible-0.075-rg178-50-ohm-coax-cable-fep-jacket-rg178b-u-p.aspx}{Pasternack RG178}}. The SiPMs are biased in pairs (one top paired with one bottom) with a low-pass RC filter ($1.5~\mathrm{k}\Omega$, $1~\mu\mathrm{F}$) in between the power supply and SiPM pair. For the results presented in this work, only three of the four SiPMs in each array were functional (six SiPMs in total). The small number of channels limits transverse position reconstruction. Each SiPM is read out individually, terminating directly into a digitizer\footnote{\href{https://www.caen.it/products/dt5730/}{CAEN DT5730B}} with a 500~MS/s sampling rate. Data acquisition is triggered when any channel exceeds a threshold of 3~mV above baseline. Event windows are 25~$\mu$s in duration, centered on the trigger.

\subsection{Cryostat and temperature control} The TPC is housed in a stainless steel inner cryostat vessel (ICV) with 7.29~cm inner diameter. When full and operational, the ICV contains between 500-700~g of xenon, depending on phase (liquid vs crystal) and height in the extraction region. The ICV sits in the outer cryostat vessel (OCV), which remains at vacuum during operation, to minimize  convective heat leak from the room temperature to the ICV. Multi-layer insulation wraps the ICV to block most of the radiative heat. The OCV is surrounded by a 5~cm-thick layer of lead covering all sides except the top to reduce backgrounds from both cosmic and nearby external radiation. 

Cooling power is supplied by a cryocooler\footnote{\href{https://www.arscryo.com/orca-mr-specifications}{Advanced Research Systems MRGS-110}} with a maximum cooling capacity of roughly 25~W near the xenon triple point at -112$^\circ$C. The copper cryocooler coldhead is connected via multiple brazed copper braids to copper clamps, one radially surrounding the bottom of the ICV and another at the top ICV flange. Great care was taken to ensure good thermal connections at both locations, to achieve azimuthal and radial temperature uniformity of the ICV while enabling the vertical temperature gradient required for the crystal formation procedure described in detail below.

To avoid hot spots that may lead to non-uniform crystal growth or local regions of liquid xenon, copper thermal connections are included in several key places (see Fig.~\ref{fig:schematic}). A copper ring mounts directly to the cold ICV top flange and, in turn, three copper rods hanging from the ring hold the TPC in place via the PTFE segments. These rods are additionally coupled to the copper plating of the top SiPM board via aluminum set screws. Heat sink braids couple the same copper rods to the ground braid of the coaxial high-voltage cables, which run up the ICV to the room temperature region of the xenon space and are typically the warmest part of the ICV. The SiPM bias and signal cables are small in diameter (0.13~cm ground braid) to minimize thermal conduction and their ground braids are cable-tied to the copper rods for further heat sinking.

Because temperature control is paramount for uniform crystal growth, resistance thermal devices (RTDs) are placed in several locations to monitor the thermal profile over time. These include the bottom outside of the ICV (typically the coldest part of the system aside from the cryohead), the outside center of the top flange of the ICV, the high-voltage cable making the cathode connection (in the xenon vapor), and the copper thermal rods from which the TPC hangs.

\subsection{Xenon handling} The ICV connects via a vertical column to a gas circulation panel at room temperature. A pump\footnote{\href{https://knf.com/fileadmin/Local_files/USA/Downloads/OEM_Process_downloads/datasheets/Datasheet_UN_035_KNF_USA.pdf}{KNF UN035}} circulates xenon vapor through a closed loop, where it is chemically purified by a hot metal getter\footnote{\href{https://puregasproducts.com/ps3.htm}{SAES MonoTorr PS3-MT3-R/N}}. Both the inlet and outlet of the circulation loop are in the vapor portion of the apparatus to allow its circulation even while the detector is operating in the crystal phase. This is in contrast to traditional liquid xenon TPCs which condense the purified xenon and send it directly to the bottom of the TPC for better mixing. Nevertheless, we have not observed evidence of purity concerns during normal operation. In between data-taking runs, the xenon is stored in two interchangeable high-pressure bottles connected to the circulation panel. At the start of a run, the xenon is filled from the high-pressure bottles into the cold, low pressure ICV, passing through the getter for initial purification. 

\subsection{Calibration sources }\label{sec:sources} To characterize the detector response to gamma and alpha interactions in both the liquid and crystalline phases, we deploy two calibration sources. First, we use a $^{210}\mathrm{Po}$ source of 5.3~MeV alphas plated onto the cathode wires near the radial center. The fixed position of the source removes possible systematics from uncertainties in reconstructed position (particularly in $(x, y)$ where the four top SiPMs are insufficient to provide significant constraints), and provides a clear calibration for the maximum drift time of the TPC. The high energy of the alphas makes this source readily distinguishable from other signals in the detector. 

For gamma calibration, we use an external $^{57}\mathrm{Co}$ source of 122 and 136~keV gammas ($\sim$88\% and $\sim$12\% branching ratios, respectively), from which photoelectric absorption produces a single merged peak in the S1 response. To minimize possible systematics from differing $(x, y)$ position distributions related to stronger attenuation in the denser crystal phase, the source is placed above the center of the TPC, and is sufficiently far above the top SiPM array such that gammas entering the TPC are nearly vertically collimated. This collimation ensures that the attenuation of $^{57}\mathrm{Co}$ gammas is independent of $(x, y)$ position. We can then apply the same narrow (smaller than the $\sim$0.4~cm attenuation length) cut on $z$ position in both liquid and crystal phases to avoid any effects of position dependence.

\section{Operation and crystal formation procedure}
\label{sec:procedure}
\subsection{Operation} Prior to filling the TPC with liquid xenon, two procedures are performed to reduce contamination of the xenon due to material outgassing, especially from the PTFE parts. First, the ICV is baked at $\sim$100$^\circ$C while being continuously pumped on for approximately 12 hours. The ICV is then filled with $\sim$2~bar of xenon vapor, which is circulated through the getter for several hours, while the ICV stays at room temperature. 

Next, the cryocooler is used to cool the ICV and the instruments inside. Once the ICV temperature reaches roughly -100$^\circ$C and xenon begins to condense, we resume filling the ICV through the getter. A consistent height of liquid xenon in the ICV across runs is achieved by filling 20~g of xenon (as gauged by a mass flow meter) beyond the gate grid. Its location is determined by observing the disappearance of events with distinct S1 and S2 signals with just the cathode grid at -2.4~kV (gate and anode at ground), indicating the loss of drift and extraction between the gate and cathode. The temperature and pressure are allowed to stabilize before data is taken. 
\subsection{Crystal formation procedure}\label{sec:crystal-formation}
It is crucial to demonstrate the ability to form a good xenon crystal using a stainless steel cryostat, which is much more practical for use in a large-scale xenon detector than the technique from Ref.~\cite{Yoo:2014dca} and related crystalline xenon work, which used a glass ICV cooled via liquid nitrogen vapor. In particular, the much lower thermal conductivity of glass allows maintenance of a larger thermal gradient, which could in principle be beneficial in the technique used. We find that this is not necessary; crystal quality is discussed further in Sec.~\ref{sec:transparency}.

\begin{figure}
\centering
\includegraphics[width=.98\textwidth,angle=0]{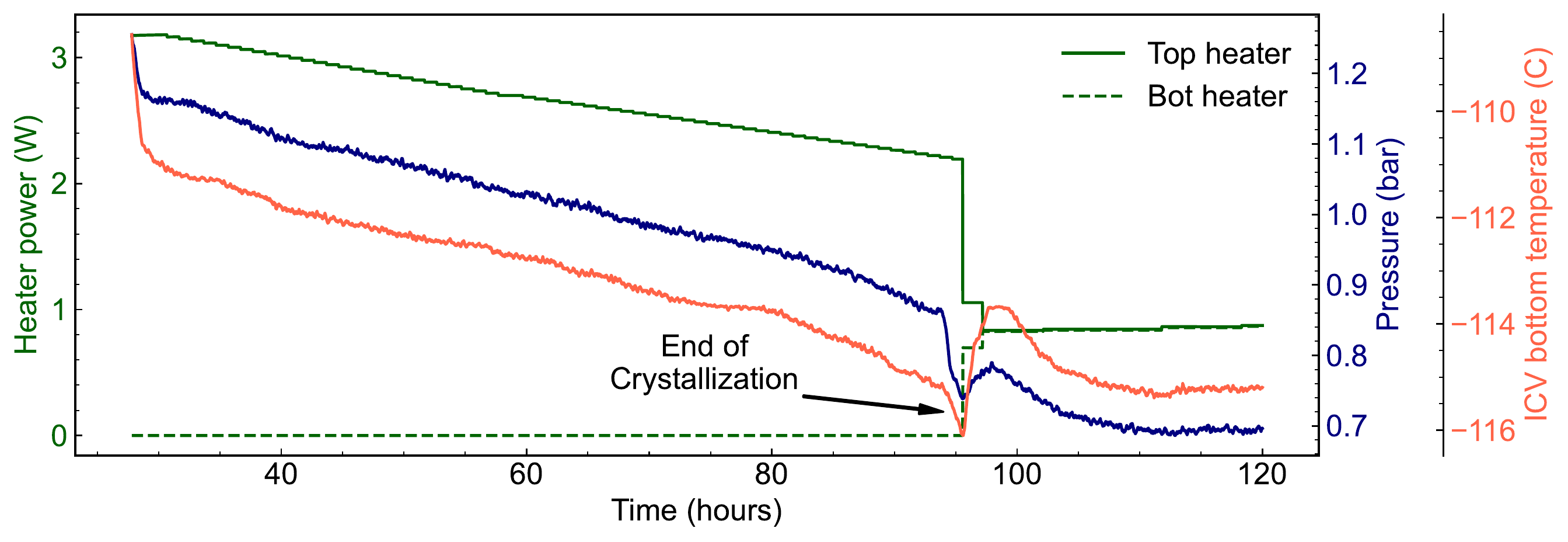}
\caption{\label{fig:freeze} 
Time evolution of xenon vapor pressure and ICV bottom temperature during crystallization. Also shown are the instantaneous power delivered by the top and bottom heaters. Heater settings are chosen to achieve an approximately constant cooling rate over time, while attempting to quickly stabilize temperatures near the triple point soon after crystallization finishes.}
\end{figure}

Before crystallization, an additional 17\% of the liquid phase mass is added to account for the higher density in crystal phase \cite{NIST, Eatwell:1961}, so that the crystal/vapor interface remains between the gate and anode grids. It has been observed, both in this work and in Ref.~\cite{Yoo:2014dca}, that crystallizing too quickly can lead to poor transparency to xenon scintillation light, perhaps due to formation of a polycrystalline structure with many domain boundaries. To avoid this, our intention is to grow the xenon crystal vertically, from the bottom of the ICV upward, following the modified Bridgeman's technique described in Ref.~\cite{Yoo:2014dca}. This approach suggests a temperature gradient of 1-2$^\circ$C/cm between the top and bottom of the ICV is preferred. We turn off the heater located on the ICV bottom during crystal formation, while the top heater on the ICV flange is used to control the process. The top heater is set to a relatively high power initially for a larger temperature gradient. As the crystal grows, the power of the top heater is decreased gradually in order to steadily cool the xenon and shift the phase boundary upward. Fig.~\ref{fig:freeze} shows a typical time evolution of the xenon vapor pressure, the bottom ICV temperature, and the instantaneous power delivered by the top and bottom vessel heaters following the procedure found to achieve good crystal quality. As the top layer of liquid xenon crystallizes, the pressure drops abruptly below the triple point at 0.818~bar. Heater power is increased to maintain the system's temperature at $\sim$-114$^\circ$C. Data collection begins once the temperature and pressure are stable. 

\section{Event reconstruction and SiPM calibration}
\label{sec:data}

\begin{figure}
\centering
\includegraphics[width=.99\textwidth,angle=0]{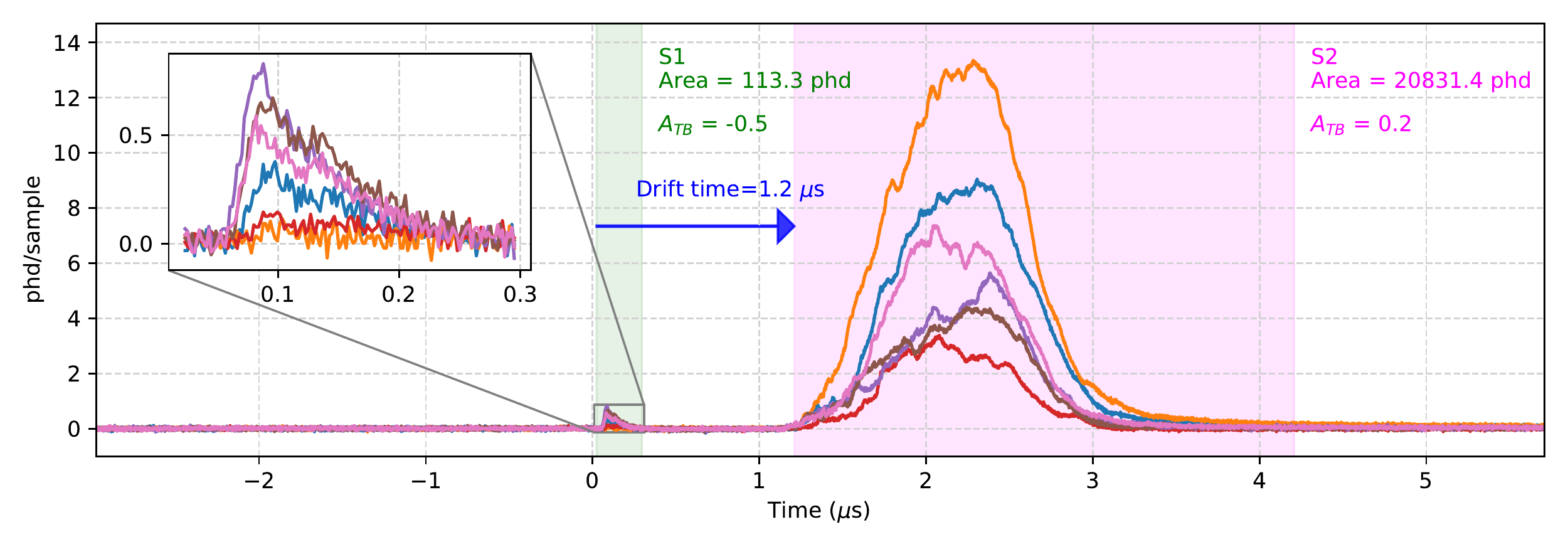}
\caption{\label{fig:waveform} A typical waveform obtained in the crystalline/vapor phase at $T=-114^\circ$C, $p=0.80$~bar, and 270~V/cm drift field, with the detector exposed to a $^{57}\mathrm{Co}$ source. The prompt, primary scintillation (S1) is followed by secondary scintillation (S2) a little over a microsecond later. The delay corresponds to the drift time of the electron cloud through the crystal, and the amplified S2 signal is caused by electroluminescence in the vapor phase after the electrons are emitted from the crystal. Each curve corresponds to a different SiPM channel after baseline subtraction. Pulse boundaries and classifications as determined by the analysis software are shown by the shaded regions, green for S1 and pink for S2, with measured pulse area summed across all channels and asymmetry in the top/bottom SiPM array ($\mathrm{A_{TB}}$) indicated, as well as the event-level drift time. The inset shows the S1 pulse in more detail.}
\end{figure}

Collected waveforms are passed through analysis software which performs baseline-subtraction, identifies pulse start and end times, classifies pulses (S1, S2, or other), and calculates summary variables both at the pulse level (e.g., pulse area) or at the event level (e.g., drift time in events with a single S1 and single S2).
An annotated example waveform from a crystalline/vapor run is shown in Fig.~\ref{fig:waveform}. Pulse variables such as top-bottom asymmetry, $\mathrm{A_{TB}} = \frac{T-B}{T+B}$, where $T$ and $B$ are the detected photon areas in the top and bottom SiPM arrays, respectively, and rise time are useful for classifying pulses as S1s (bottom-focused, sharp rise) or S2s (top-focused, slow rise).

\begin{figure}
\centering
\includegraphics[width=.49\textwidth,angle=0]{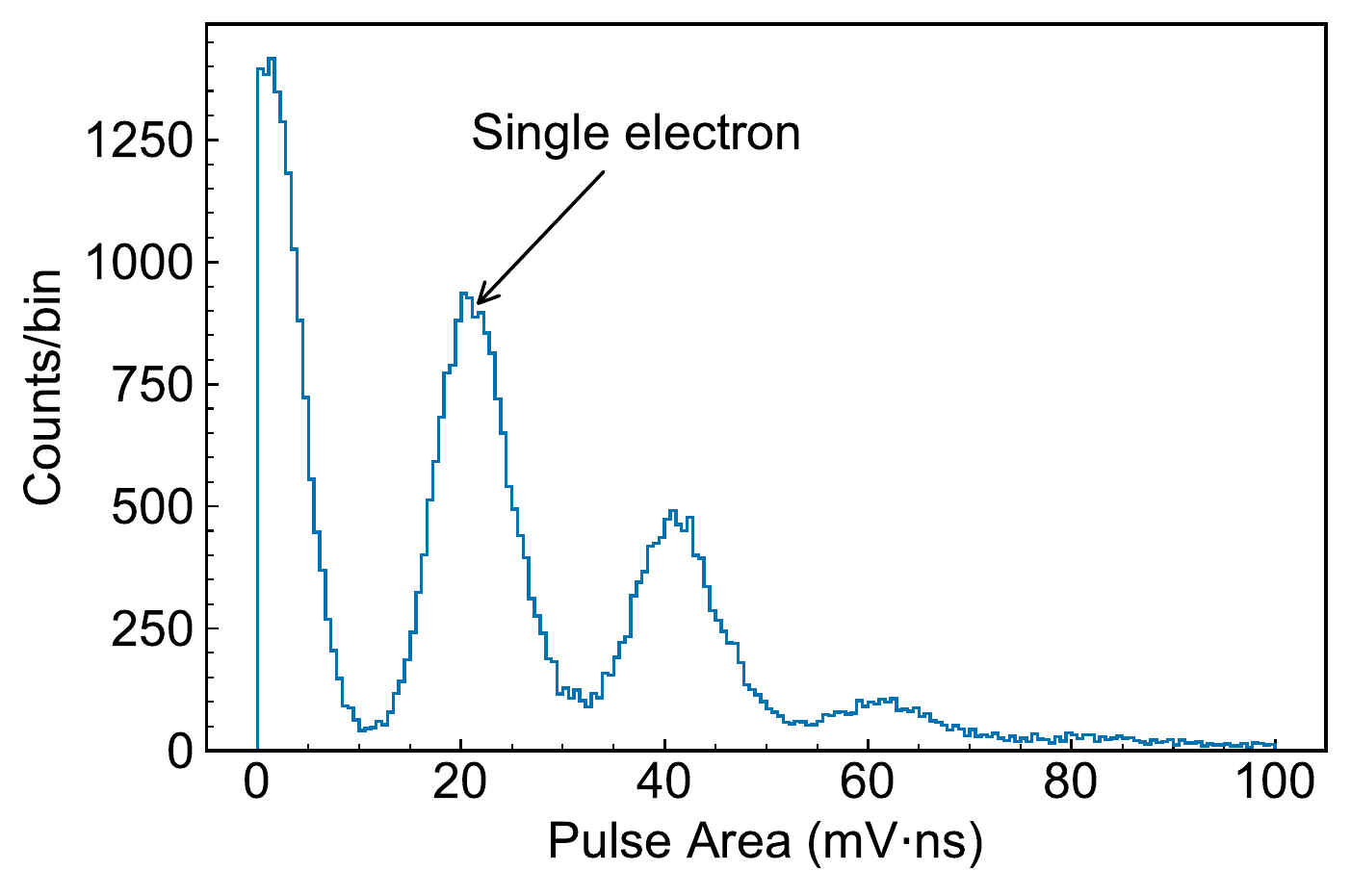}
\includegraphics[width=.49\textwidth,angle=0]{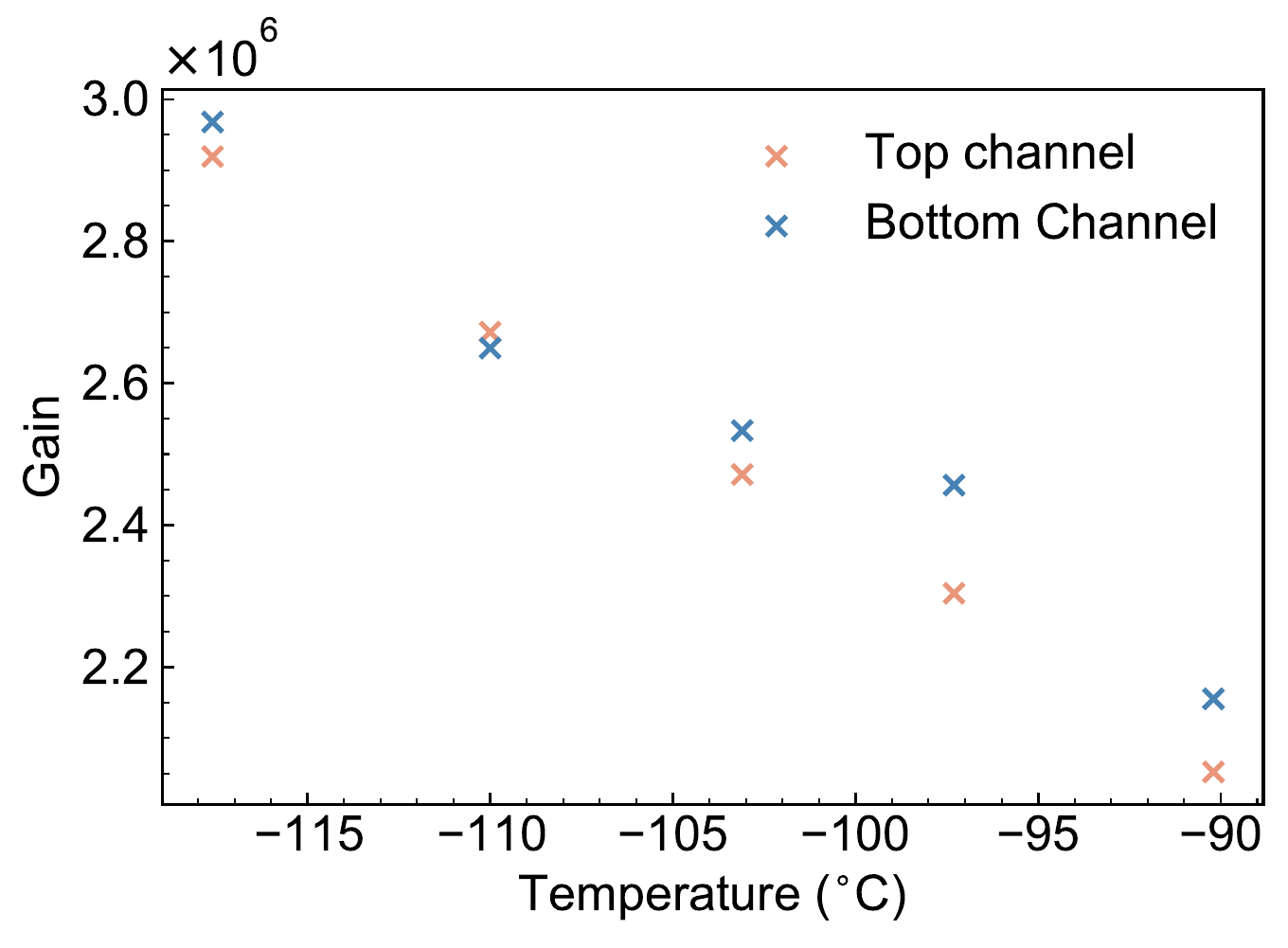}
\caption{\label{fig:calib}{\bf Left:} 
Typical spectrum of pulse area collected from SiPM dark counts (no xenon present) at $T=-114^\circ$C. Peaks from successive numbers of firing SiPM pixels are observed. A Gaussian fit to the single electron avalanche peak is used to extract the mean single photon pulse area.
{\bf Right:} SiPM gain as a function of temperature measured for a top and bottom SiPM channel, measured in a vacuum environment.}
\end{figure}

We calibrate the SiPM single photon response to provide an absolute conversion between pulse area, in units of mV$\cdot$ns, to number of photons detected (phd). During this calibration, no light is deliberately introduced to the SiPMs; rather, the trigger threshold is set close to the baseline to ensure dark counts (avalanches triggered by thermal excitations rather than photons) from the SiPM pixels dominate. The pulse areas of dark count waveforms form a clear series of peaks, shown in Fig.~\ref{fig:calib}, left, each of which corresponds to a different integer number of pixels firing. The statistical error in this step is estimated to be 2\%. We calculate the SiPM gain as $G = \mathrm{SPE}/(R\,e)$, where SPE is the pulse area of a single photon determined by a fit to the spectrum of collected dark counts (e.g., Fig.~\ref{fig:calib}, left) for each sensor, $R$ is the 50~$\Omega$ resistance of the readout circuit, and $e$ is the charge of the electron.  
For accurate comparisons between liquid and crystal phase data, the temperature-dependent SiPM gain for each channel was characterized. Fig.~\ref{fig:calib}, right indicates a change in gain of $\sim$ 1\%/$^\circ$C. The gain values used to process data are taken in liquid and crystal xenon separately, to correct this temperature dependence. During data taking, the system temperature fluctuates by $\pm0.7^\circ$C, introducing an error of $\pm0.7$\%. 

The effect of optical cross talk, in which the avalanche in a SiPM pixel may emit a secondary photon and induce an avalanche in a nearby pixel, is also considered. This effect does not change the size of the single pixel response determined from the calibration described above, but does increase the area of pulses from interactions in the xenon by the same percentage as its probability of occurrence. This probability depends on factors such as the SiPM's gain and the distance between pixels, and is of order 5\% for the SiPMs and gains used here \cite{Baudis:2018pdv, nEXO:2018cev,Baudis:2021dsq}. To fairly compare pulse size between the crystal and liquid states, we adjust the SiPMs' bias when in the crystalline state to ensure the same value of SPE as in liquid.

Especially large pulses can exhibit saturation effects when individual pixels within each SiPM device begin to have an appreciable chance of absorbing multiple photons within the single pixel reset time. Assuming uniform illumination of the eight SiPMs (each with 6400 pixels), saturation begins to have a >10\% effect for pulse areas greater than $\approx2\times10^4$~phd within a single reset time (as is the case for the narrow S1 pulse)~\cite{Grodzicka:2015}.
S2s can show a smaller effect for the same pulse area due to their significantly broader photon arrival time profile, which allow some pixels to fire multiple times within a single pulse~\cite{Vinogradov:2011}. In practice, we do not see evidence of saturation for S1 signals from either the $^{57}\mathrm{Co}$ or $^{210}\mathrm{Po}$ calibration sources, though the $^{210}\mathrm{Po}$ S2s are large enough to produce a clear saturation effect (the pulse area shows a sharp falling edge, and increasing the S2 gain through higher extraction field has a minimal effect on S2 area). In the following results, we only use the $^{210}\mathrm{Po}$ S2 signal to define the maximum drift time of the TPC, and do not make use of the estimated pulse area.

\section{Results}

\subsection{Scintillation in crystalline vs liquid xenon}
\label{sec:scint}

In Sec.~\ref{sec:data} we established the capability of the crystalline TPC to function fully as intended, i.e., to reliably observe both S1 and S2 signals. In this section, we quantitatively compare the S1 properties between crystal and liquid xenon for both gamma and alpha sources, demonstrating approximately equal sensitivity to the scintillation channel in both phases. 

Unless otherwise noted, results are averaged across a fiducial drift region to avoid electric field fringing effects near the cathode and gate, and are taken well after the crystal formation has finished to avoid potential effects as temperatures in the crystal stabilize. This drift time cut is intended to select the same vertical section of the TPC in both phases, spanning 1.5-2.5~$\mu$s in crystal phase and 3.6-5.0~$\mu$s in liquid. Conversion between drift time and vertical position is done using the known drift region extent of 0.74~cm, the maximum drift time measured using $^{210}\mathrm{Po}$ decays on the cathode, and the drift time of the gate, estimated using features in the data such as S1 size versus drift time, which drops when the field increases above the gate due to reduced electron-ion recombination. Reference liquid data is taken at $T=-106^\circ$C and $p=1.20$~bar, while crystal phase data after stabilization is obtained at $T=-114^\circ$C and $p=0.80$~bar. 

\begin{figure}
\centering
\includegraphics[width=.49\textwidth,angle=0]{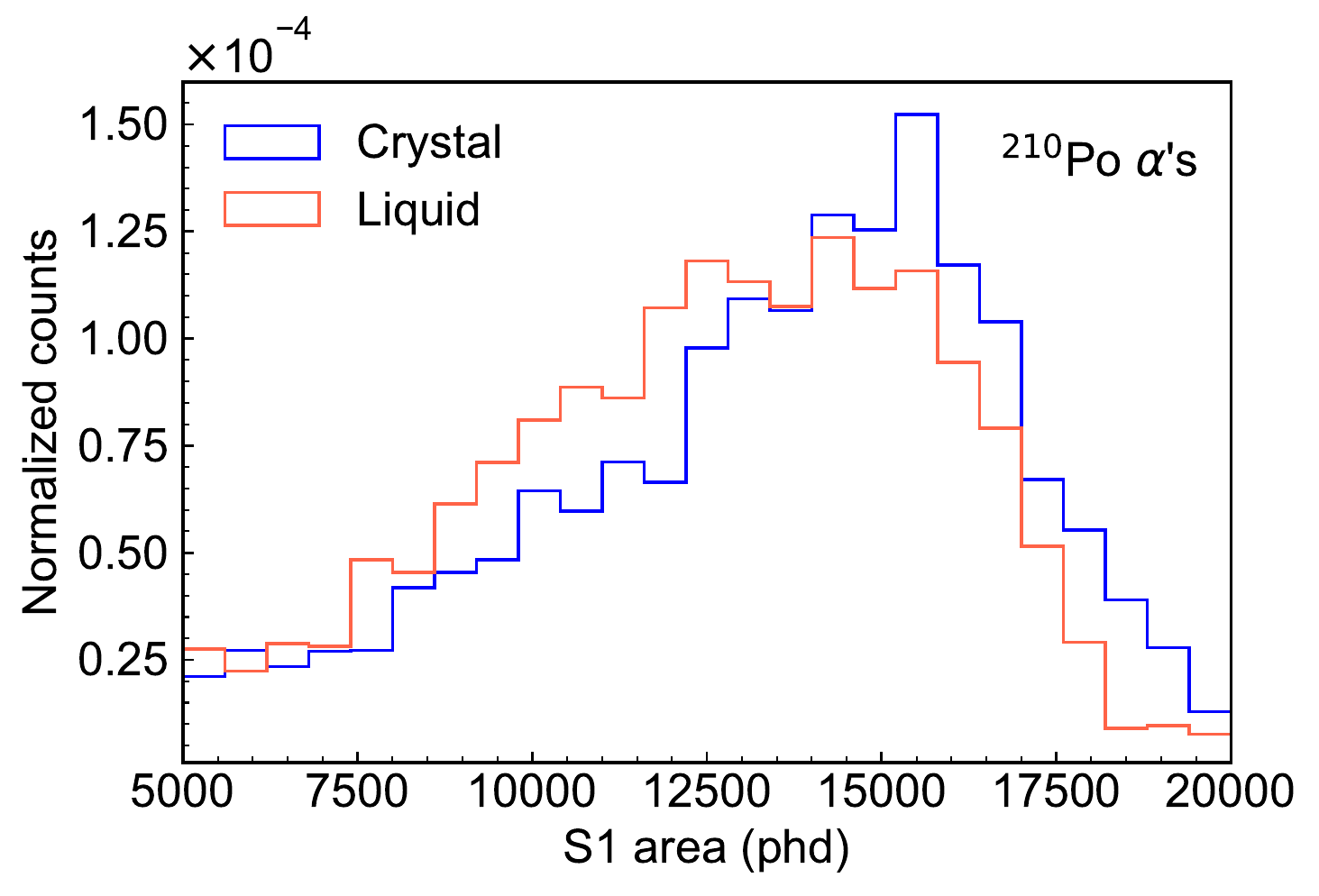}
\includegraphics[width=.49\textwidth,angle=0]{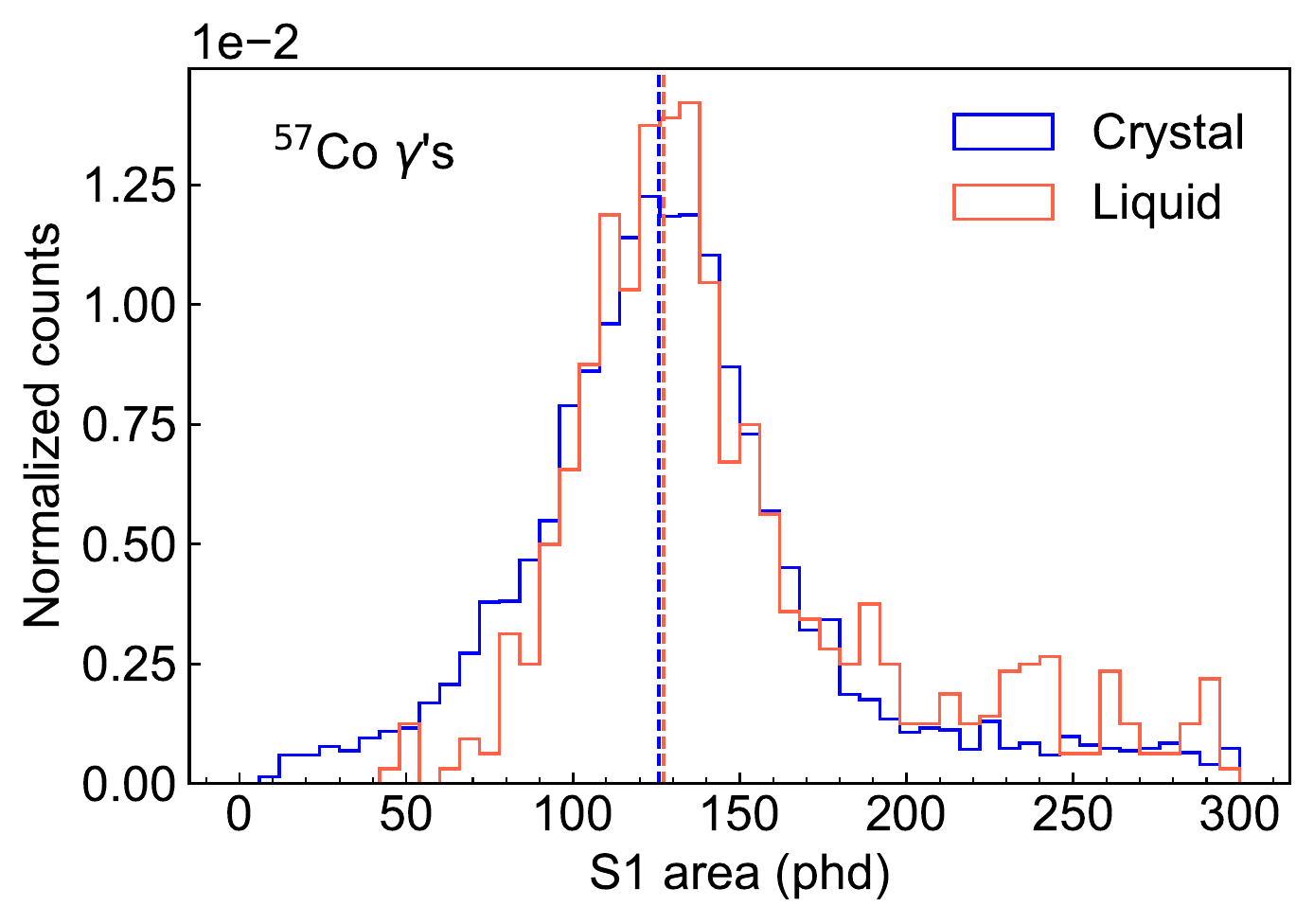}
\caption{\label{fig:po-co-s1} Observed scintillation spectra in liquid and crystalline xenon, in units of photons detected (phd). \textbf{Left:} $^{210}\mathrm{Po}$ spectra with cathode and gate set to 0~V. The non-Gaussian shape is caused by shadowing from the cathode wires, as explained in Sec.~\ref{sec:transparency}. \textbf{Right:} $^{57}\mathrm{Co}$ spectra with the nominal 270~V/cm drift field, averaged across the fiducial drift region. Mean value is indicated by the dashed lines; uncertainty on the mean (from Gaussian fit) is approximately 2~phd.}
\end{figure}

Fig.~\ref{fig:po-co-s1}, left compares the S1 spectra from the $^{210}\mathrm{Po}$ alpha source on the cathode obtained in the liquid and crystal phase with the cathode and gate set to 0~V, chosen to avoid ambiguity in the field strength close to the cathode wires where it varies rapidly with distance from the wire surface.
We find a nearly identical scintillation response of liquid and crystalline xenon, with a possible slight increase for alpha particles in the crystalline case. Fig.~\ref{fig:po-co-s1}, right provides the equivalent comparison for the $^{57}\mathrm{Co}$ external gamma source with the cathode at -2.6~kV and the gate at -2.4~kV, producing an average drift field of 270~V/cm. There is good agreement between the two spectra, with no evidence for the 15\% reduction in S1 yield in the crystal phase observed in Ref.~\cite{Yoo:2014dca}. We suspect this result is due to our choice of photosensor (SiPM) compared with the traditional PMT used in Ref. \cite{Yoo:2014dca}. This can be explained by the known component of crystalline xenon scintillation which is shifted to a wavelength as small as 148~nm \cite{Varding:1994abc}. In contrast with the usual liquid/gaseous xenon scintillation wavelength of 175~nm, this crystalline component is expected to be severely attenuated by most window materials. For example, the quartz glass used in many VUV-sensitive photomultipliers has a cut-off around 160~nm.

\begin{figure}
\centering
\includegraphics[width=.49\textwidth,angle=0]{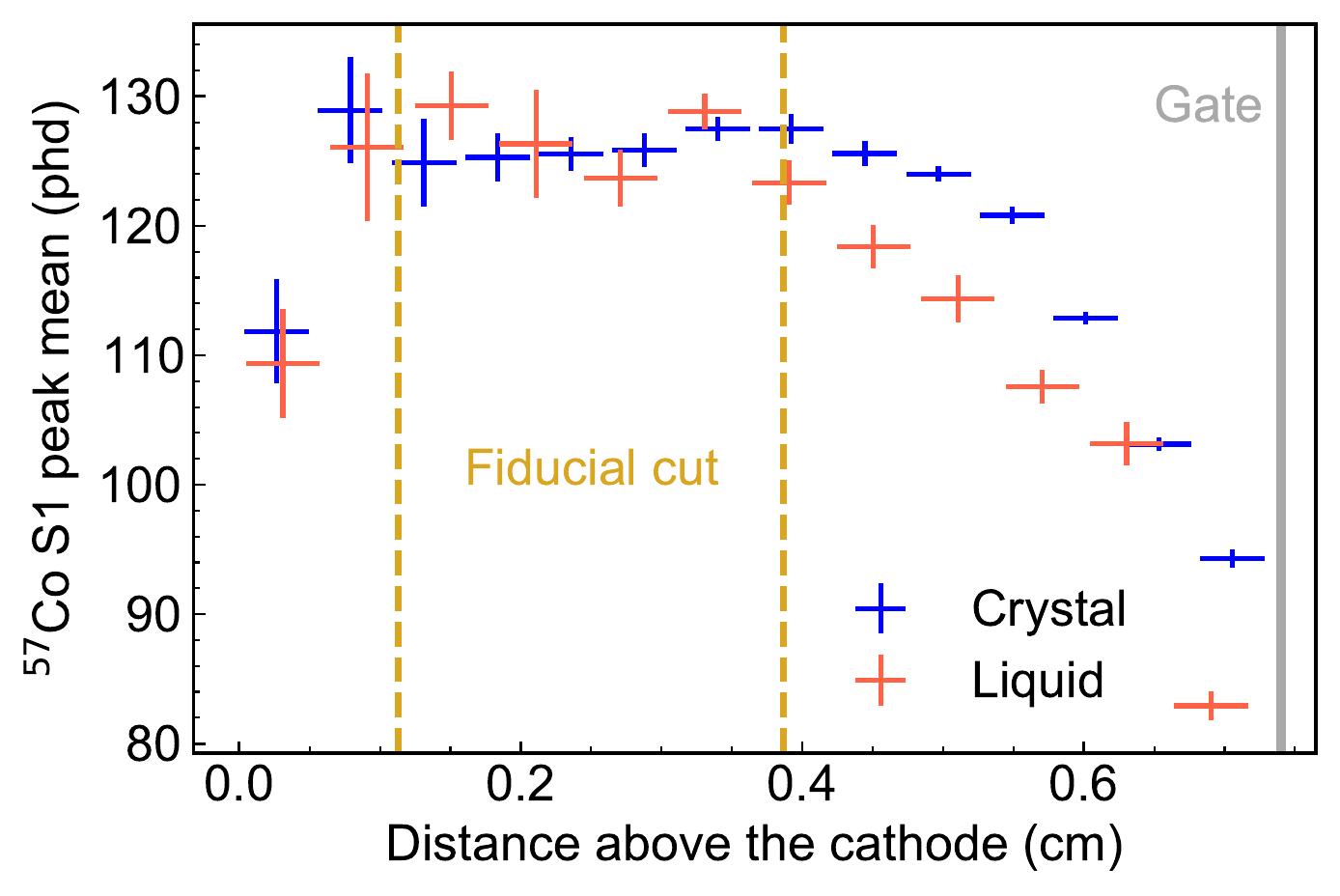}
\includegraphics[width=.49\textwidth,angle=0]{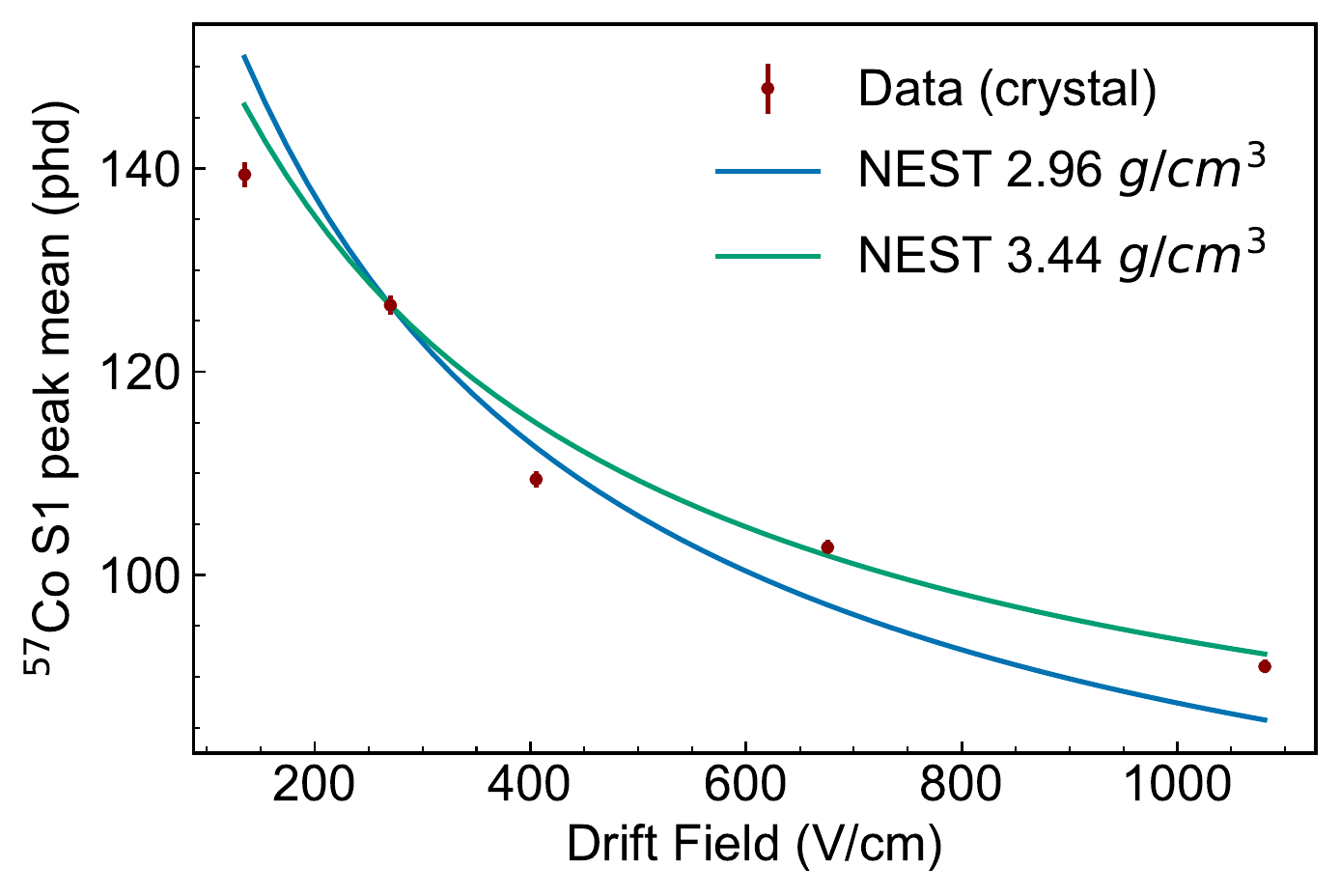}
\caption{\label{fig:co-s1-drift} \textbf{Left:} Scintillation signal at different locations in liquid and crystalline xenon with 270~V/cm drift field, in units of photons detected (phd) for $^{57}\mathrm{Co}$. The fiducial drift cut used for position-averaged S1 spectra is defined by the dashed lines, and is intended to avoid regions of electric field non-uniformity. \textbf{Right:} $^{57}\mathrm{Co}$ S1 size, averaged across the fiducial drift region, vs nominal electric field. NEST values represent expectations for field dependence in liquid xenon and are scaled to match the measurements at the typical operating field of 270~V/cm to emphasize only relative changes. The two xenon densities used are representative for liquid (2.96~g/cm$^3$) and crystal (3.44~g/cm$^3$) phase.}
\end{figure}

Fig.~\ref{fig:co-s1-drift}, left compares the mean obtained in a Gaussian fit to the $^{57}\mathrm{Co}$ S1 spectrum, as a function of vertical position, between liquid and crystal phase at 270~V/cm average drift field. Here again we see good agreement ($<10\%$ discrepancy) at all positions, aside from a slight discrepancy near the gate position where the electric field is sharply changing. The decreases in S1 size near the cathode and in the extraction region in both phases are likely to be due to the higher electric field, which causes fewer electrons to recombine and produce S1 photons.

Fig.~\ref{fig:co-s1-drift}, right shows the dependence of the mean $^{57}\mathrm{Co}$ S1 size in the crystal phase on the nominal electric field in the drift region. As expected, S1 size decreases with increasing electric field; the field dependence observed is roughly consistent with expectations from Noble Element Simulation Technique (NEST) v2.3.0~\cite{NESTv2} simulations in liquid xenon, suggesting similar recombination properties for electron recoil events in both crystal and liquid. The precise field may vary from the nominal value (parallel plate approximation) due to fringing near the TPC walls and grid wires, though this effect is reduced by the fiducial drift region selection.

\subsection{Crystal transparency}\label{sec:transparency}
This section discusses evidence that the crystal formed by the procedure described in Sec.~\ref{sec:crystal-formation} is highly transparent, using the relative size of S1 signals following different light propagation paths.

\begin{figure}
\centering
\includegraphics[width=.55\textwidth,angle=0]{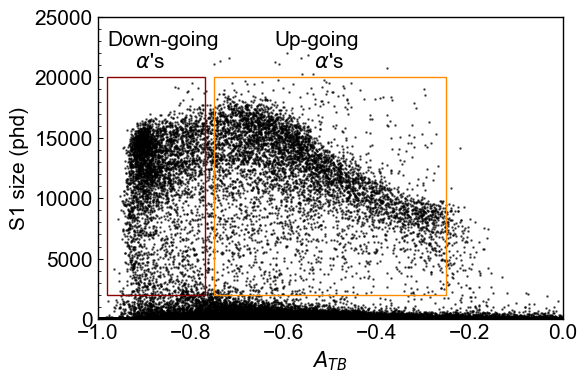}
\includegraphics[width=.38\textwidth,angle=0]{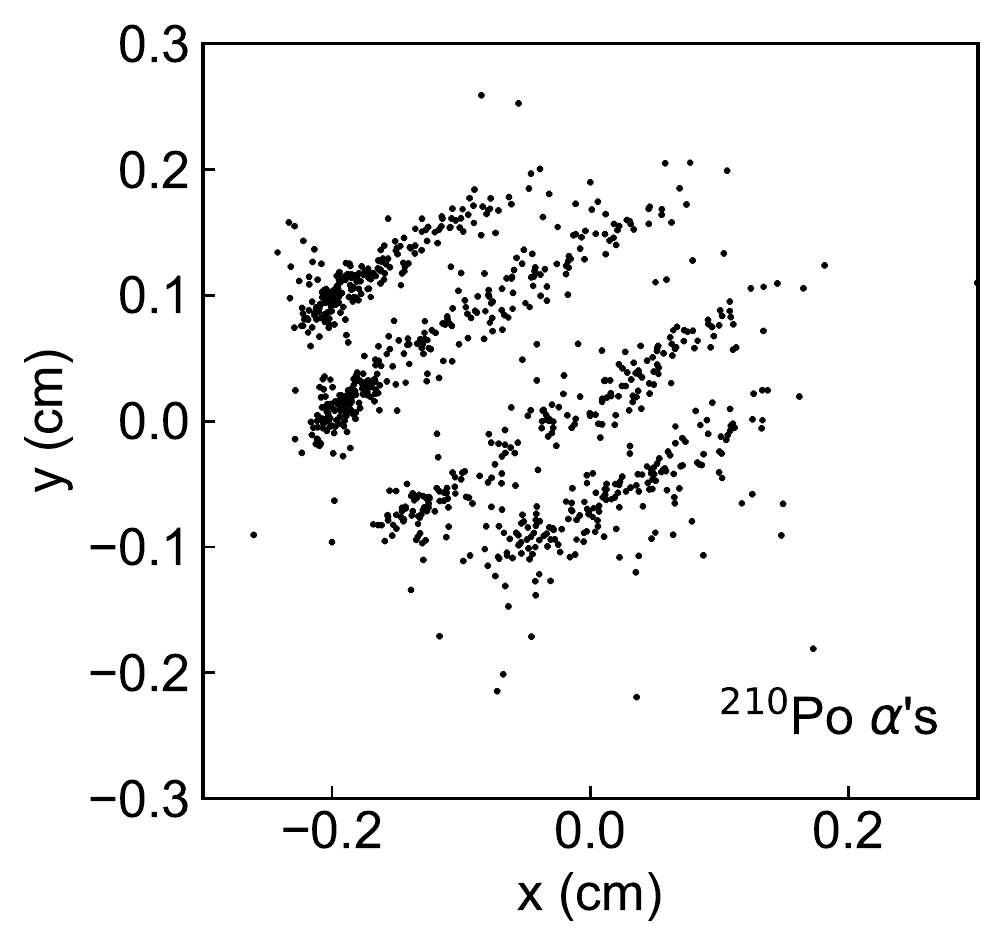}
\caption{\label{fig:po-shadow}\textbf{Left:}  Scatter plot of S1 pulse area versus top-bottom asymmetry, focused on the region of interest for $^{210}\mathrm{Po}$ events which originate from the cathode. The cuts which define the up-going and down-going $^{210}\mathrm{Po}$ populations are indicated. \textbf{Right:} Position distribution of down-going $^{210}\mathrm{Po}$ decays as estimated by the S1-weighted centroid of the bottom SiPM array positions, showing the positions of the cathode wires where decays originate. Position localization is only possible when the majority of the light travels directly into the SiPMs, rather than reflecting off of the surrounding PTFE or at the vapor interface.}
\end{figure}

The cathode wire diameter (100~$\mu$m) on which the $^{210}\mathrm{Po}$ source is deposited is significantly larger than the 5.3~MeV alpha range in condensed xenon ($\sim$40~$\mu$m), which leads to a strong S1 shadowing effect that depends on the position of the decay along the circumference of the wires. Fig.~\ref{fig:po-shadow} shows this shadowing in the space of S1 area versus $\mathrm{A_{TB}}$ (defined in Sec \ref{sec:data}). Pulses from decays on the bottom half of the wires form the indicated ``down-going'' population, where the cathode shadows S1 light travelling toward the top SiPM array and leads to a very small $\mathrm{A_{TB}}$; conversely, the ``up-going'' population has a larger $\mathrm{A_{TB}}$, though still negative, as much of the S1 light on an initial upward trajectory is reflected at the interface with the xenon vapor. This shadowing leads to an overall S1 spectrum that is non-Gaussian; this does not impede comparisons between liquid and crystal phase, as shadowing effects will be the same in both cases.

\begin{figure}
\centering
\includegraphics[width=.49\textwidth,angle=0]{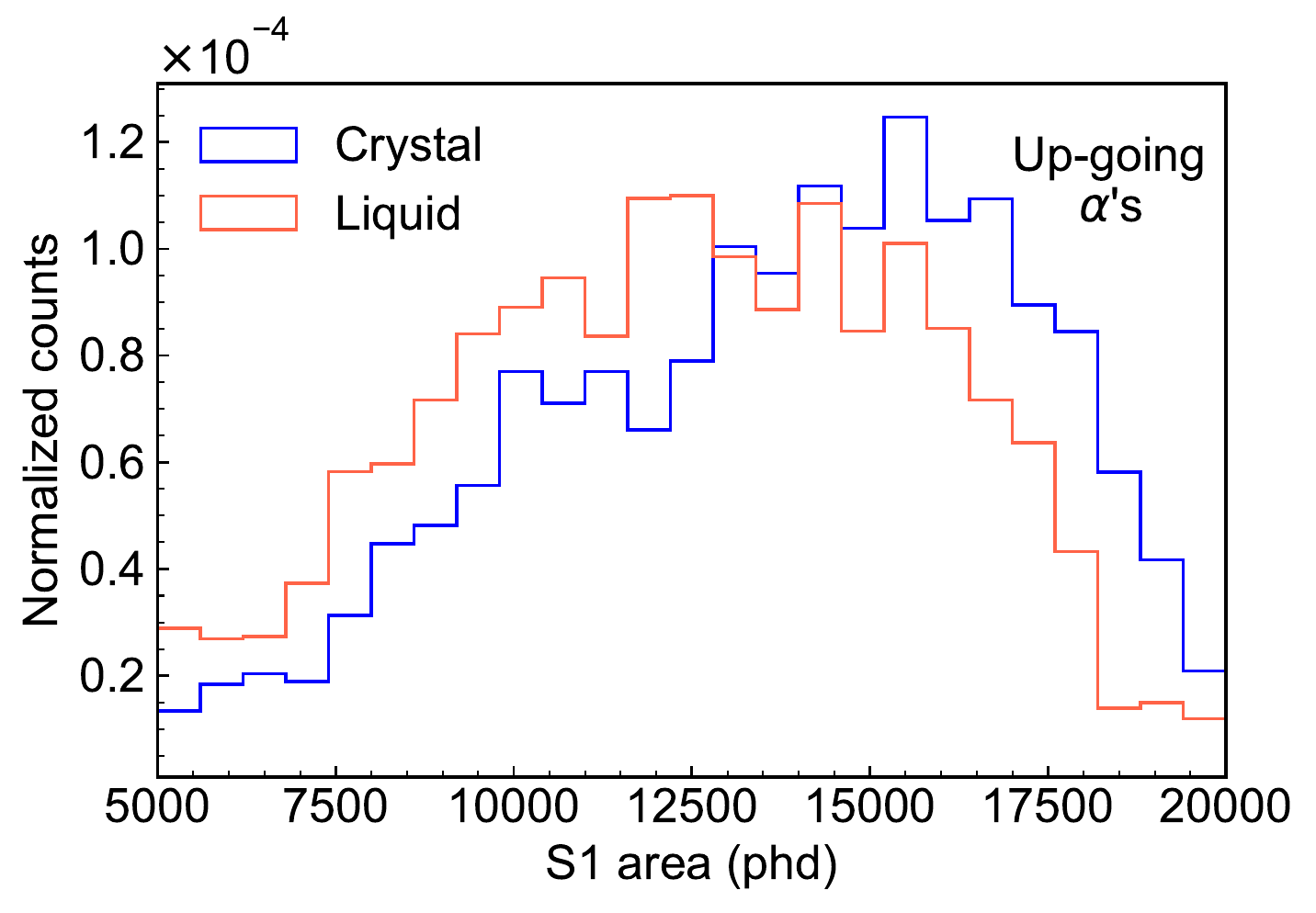}
\includegraphics[width=.49\textwidth,angle=0]{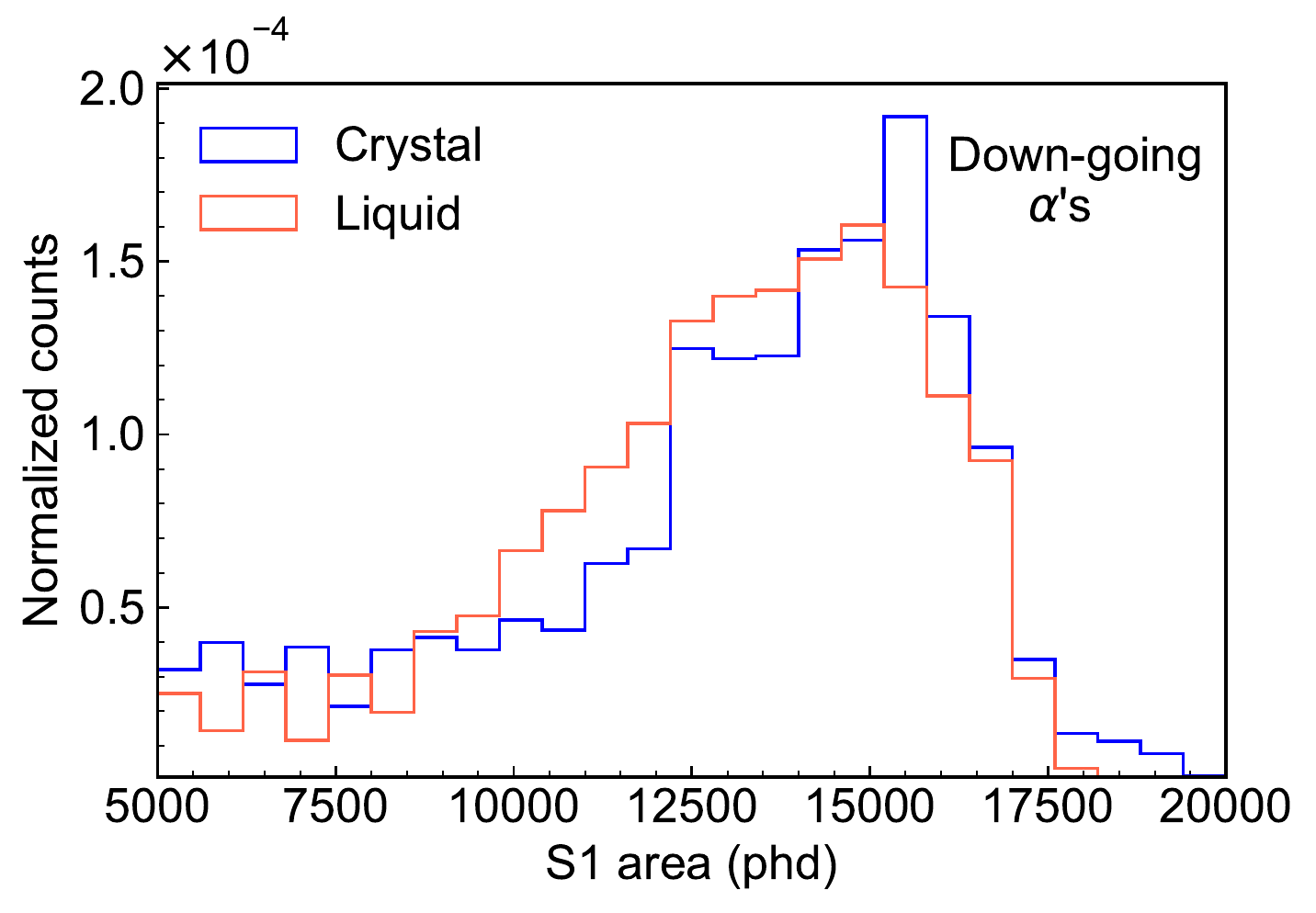}
\caption{\label{fig:po-s1-up-down} Observed scintillation spectra for $^{210}\mathrm{Po}$ with the cathode and gate set to 0~V, for decays from the top of the cathode wires (\textbf{left}) and those from the bottom (\textbf{right}). The up-going sample shows stronger evidence for a slight increase in the number of detected photons in the crystal phase, providing evidence against crystal opacity which would preferentially reduce the S1 signal from up-going light.
}
\end{figure}

The shadowing effect provides a useful handle on light propagation through the xenon volume. Fig.~\ref{fig:po-s1-up-down} shows the $^{210}\mathrm{Po}$ S1 from Fig.~\ref{fig:po-co-s1} separated into up-going and down-going components. The similarity in the liquid-to-crystal S1 ratio between up-going and down-going events is a good indicator of high crystal transparency: if the crystal were opaque to S1 light, the up-going signal would show a stronger decrease relative to the down-going signal, due to the fact that light from down-going decays primarily travels directly into the bottom SiPM array. In contrast, light from up-going events primarily travels through the crystal and reflects at the crystal/vapor interface before reaching the bottom SiPM array.

The depth dependence of the $^{57}\mathrm{Co}$ S1 peak shown in Fig.~\ref{fig:co-s1-drift} provides a second indication of good crystal transparency: if the transparency to S1 light were poor, we would expect that the S1 size would be reduced at shorter drift times (relative to that in liquid) as the S1 light is focused in the bottom SiPM array and light coming from the bottom of the detector will have less distance to travel through the crystal on average. Similar arguments provide evidence that possible changes in reflectivity at the PTFE-xenon interface do not substantially reduce the S1 size in the crystal phase as compared to liquid: since most of the light from down-going $^{210}\mathrm{Po}$ decays ($^{57}\mathrm{Co}$ near the cathode) travels directly into the bottom SiPM array without reflection, it will be affected less by reflectivity changes than up-going $^{210}\mathrm{Po}$ decays ($^{57}\mathrm{Co}$ near the gate). 

\subsection{Crystal evolution over time}

Unlike in liquid phase, crystalline xenon has internal structure that can, in principle, change over time, especially soon after freezing finishes when temperatures of the crystal are still changing and thermal contraction can affect contact of the xenon with nearby components such as SiPMs and the TPC walls. It is also crucial that results are repeatable from one run to the next, given a fixed crystallization procedure. To investigate this, we have studied the S1 size dependence over time in our calibration sources in two runs. 

\begin{figure}
\centering
\includegraphics[width=.98\textwidth,angle=0]{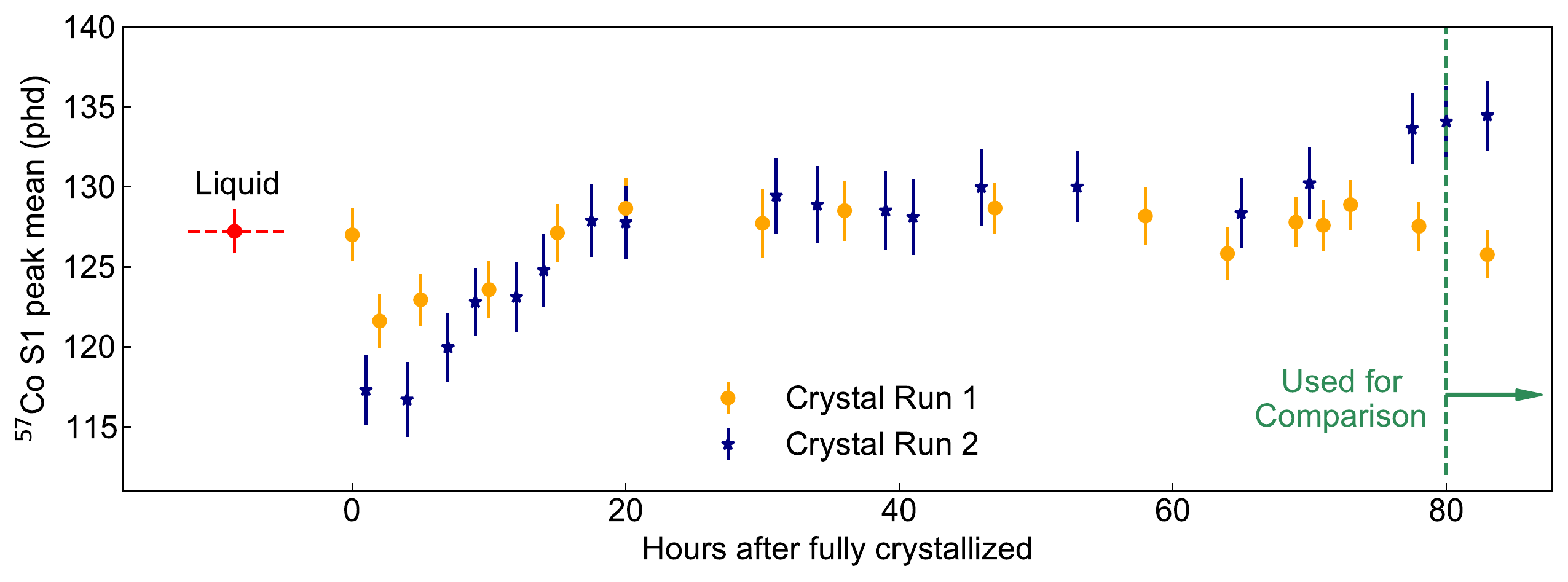}
\caption{\label{fig:co-s1-time} Mean $^{57}\mathrm{Co}$ S1 size evolution in crystal xenon with 270~V/cm drift field versus time since completion of crystallization (corresponding to $t\approx95$~hr in Fig.~\ref{fig:freeze}) for two different runs. Data taken beyond the dashed line is used for comparison with liquid phase, ensuring transient effects on S1 size are not included.}
\end{figure}

Fig.~\ref{fig:co-s1-time} shows the variation in mean $^{57}\mathrm{Co}$ S1 size with time since the completion of crystallization for two different runs using the same crystal formation procedure (described in Sec.~\ref{sec:procedure}). This indicates that following this procedure is sufficient to produce repeatable results in the crystal phase. There also appears to be a period of roughly 1~day after crystal formation finishes during which the S1 size continues to change; after this period, it appears stable. 

\subsection{Charge signals in crystal xenon}
\begin{figure}
\centering
\includegraphics[width=.49\textwidth,angle=0]{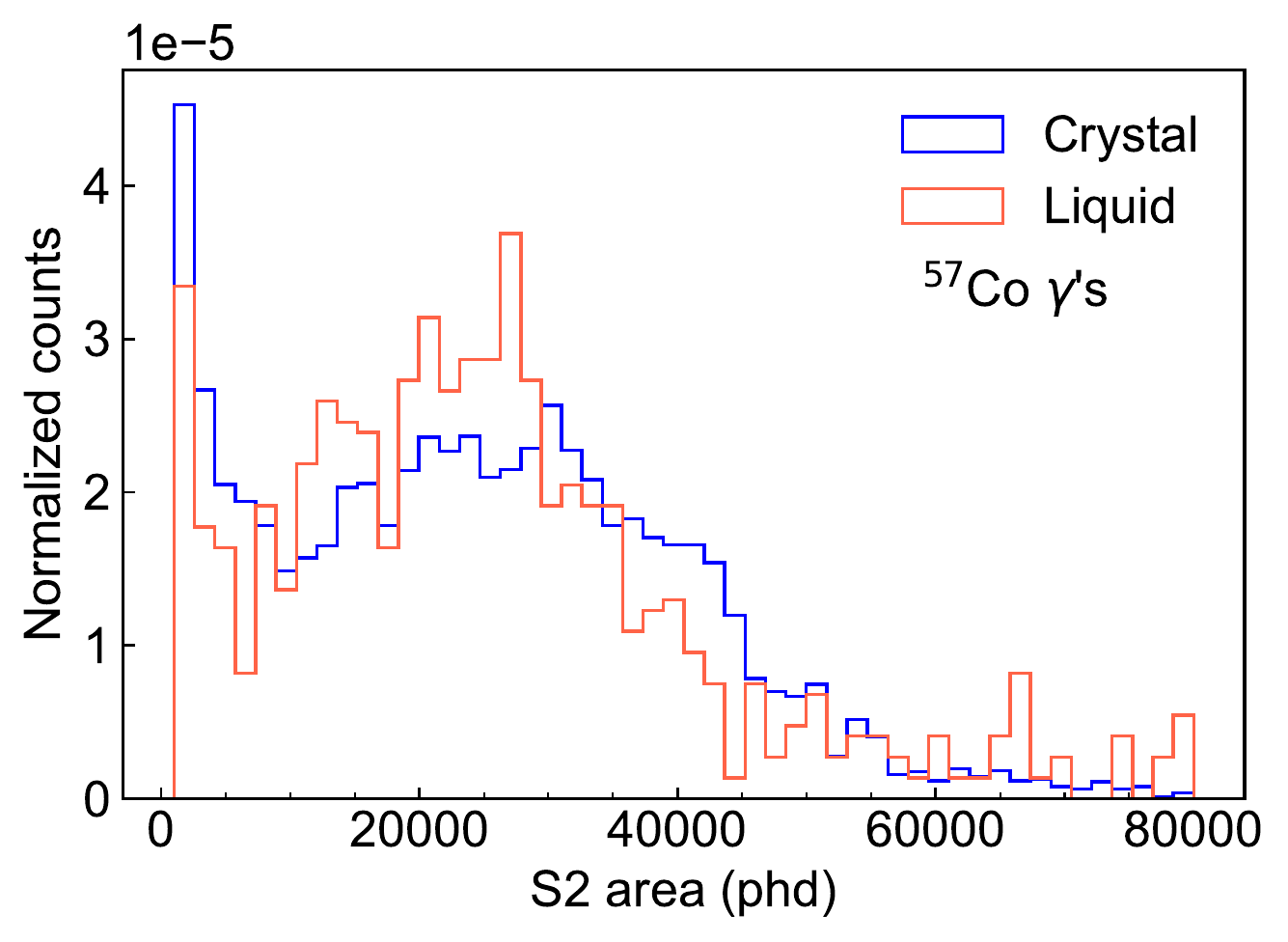}
\includegraphics[width=.49\textwidth,angle=0]{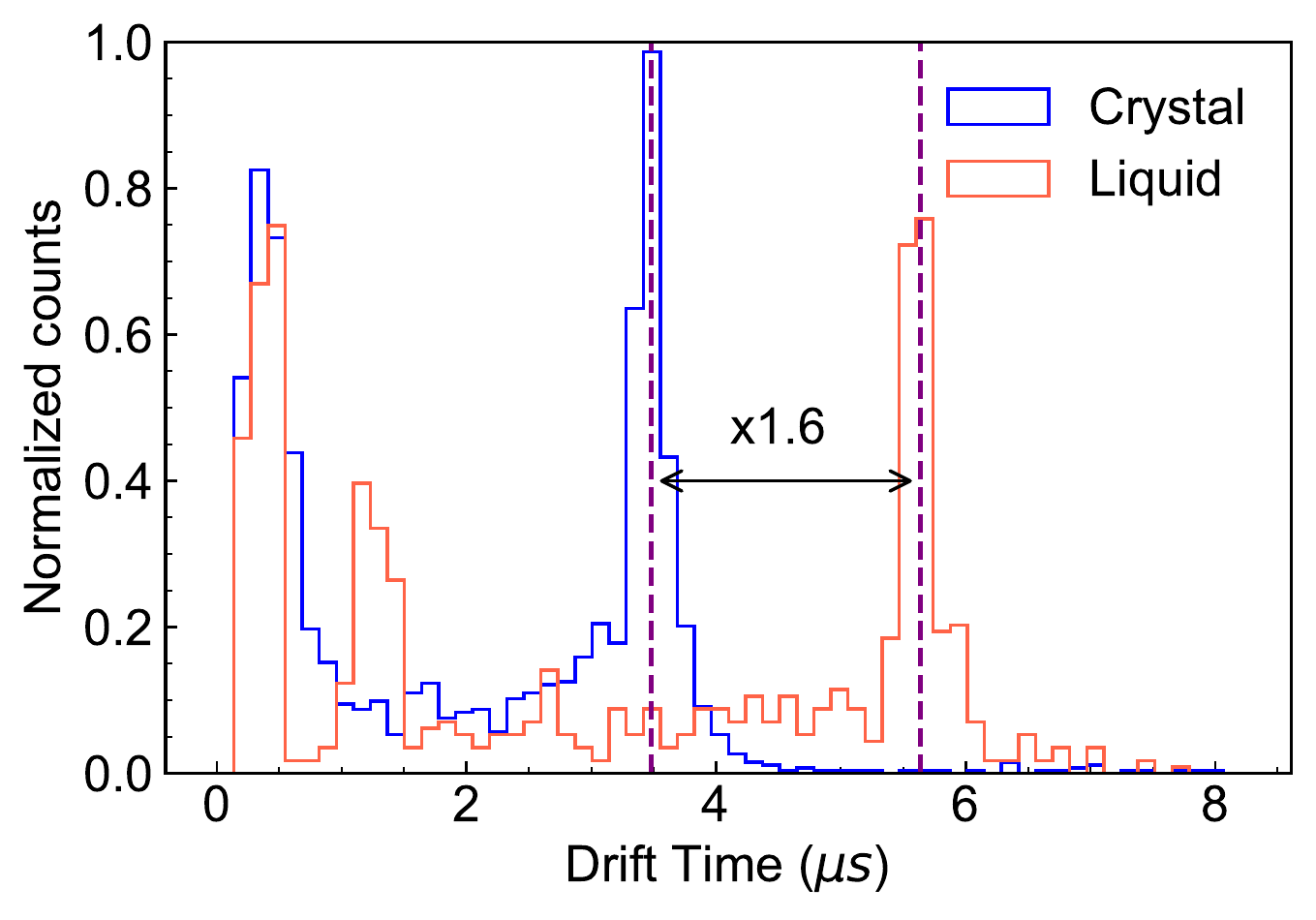}
\caption{\label{fig:co-s2}\textbf{Left:} Mean $^{57}\mathrm{Co}$ S2 size in two sample runs in liquid and crystal xenon with 270~V/cm drift field. Due to uncertainty in the location of the vapor interface, this comparison should be considered as illustrative rather than strictly quantitative. \textbf{Right:} Electron drift time distribution of $^{210}\mathrm{Po}$ alpha decays originating from the cathode wires, using the up-going $^{210}\mathrm{Po}$ selection defined in Fig.~\ref{fig:po-shadow}, for liquid/vapor and crystalline/vapor conditions. The peak from alpha decays on the cathode shows that the detector's maximum drift time decreases by a factor of 1.6 between the liquid and crystalline phases.
}
\end{figure}

Precise comparisons of the charge yield between liquid and crystal phase are challenging due to the unknown variation in liquid/crystal height, which affects S2 gain, as well as the electron extraction efficiency. In crystal phase, the crystal/vapor interface may not even be planar (e.g., freezing may occur faster near the walls), complicating estimates of both electron extraction and S2 gain. Nevertheless, we find no significant qualitative differences in the S2 signal size for the crystal phase, as indicated in Fig.~\ref{fig:co-s2}, left. This qualitative picture is consistent across multiple runs. A similar comparison for the $^{210}\mathrm{Po}$ source is not practical due to strong SiPM saturation effects for the much larger S2 signal.

The maximum electron drift time as measured by the $^{210}\mathrm{Po}$ peak is $\times1.6$ smaller in the crystal phase, as shown in Fig.~\ref{fig:co-s2} (right). This corresponds to a factor $\times1.45\pm0.15$ increase in the electron drift velocity, subject to uncertainty in the height of the vapor interface relative to the gate electrode. The effect appears to be somewhat dependent on the temperature of the crystal phase. This is consistent with expectations from Ref.~\cite{Yoo:2014dca} which indicate faster charge drift in the crystal phase and with drift speed increasing as temperature decreases. These effects are likely due to reduced electron-phonon scattering. The increase in drift speed may be beneficial for reducing backgrounds from the accidental coincidence of uncorrelated S1 and S2 signals, due to the reduction in the maximum drift time window, which grows in relative importance as xenon detectors increase in size.

\section{Conclusions}
We have demonstrated the basic particle detection capabilities of a crystalline/vapor xenon TPC, with an eye to its potential to replace liquid/vapor xenon TPCs. We have established a repeatable recipe for generating transparent, time-stable crystals in a stainless steel cryostat with a cooling system similar to that of much larger existing TPCs. We find that the scintillation signal size in the crystal phase is comparable to that in liquid, for both gamma rays and alphas, in contrast to earlier measurements. Its dependence on electric field strength also follows a similar trend to that in liquid xenon. Charge signals are reliably extracted from crystal to vapor, with no evidence for decreased size in crystal phase. Electron drift speeds are furthermore increased in the crystal for the same electric field strength, consistent with prior measurements. 

We are presently upgrading the instrument to include 32 channels of SiPMs so that we can precisely reconstruct the $(x,y)$ position of interaction vertices and increase the overall light collection of the detector. This upgrade will allow us to demonstrate the radon-tagging and exclusion capabilities described in the Introduction. Successful radon removal can enable xenon TPCs to reach a new regime of sensitivity to dark matter, with potential for further improvement from reduction of backgrounds from accidental coincidence signals and delayed electron noise. 

Future work will also investigate the potential for single electron sensitivity and measurement of delayed electron noise, enabled by increased light collection. The ability to measure the S2 size of single electrons as a function of $(x,y)$ position will consequently lead to more precise understanding of the vapor interface location and enable a direct comparison of charge yields between liquid and crystal phases. In addition, by comparing the response of SiPMs with and without a quartz window, which strongly attenuates light in the VUV, we hope to gauge the possible effect of a lower-wavelength scintillation component in the crystal phase. Lastly, we plan to perform further studies of longer-term stability of operation.

\begin{acknowledgments}
This material is based upon work supported by the U.S. Department of Energy, Office of Science, Office of High Energy Physics, under award number DE-AC02-05CH1123.
\end{acknowledgments}

\bibliographystyle{JHEP} 
\bibliography{main.bib}

\end{document}